\documentclass[lcws2002]{article}
\usepackage[dvips]{epsfig}
\usepackage{amstext,amsmath,amssymb}

\newcommand{\half}{{\textstyle\frac{1}{2}}}

\def\gsim{\mathrel{\rlap{\raise 1.5pt \hbox{$>$}}\lower 3.5pt
\hbox{$\sim$}}}
\def\lsim{\mathrel{\rlap{\raise 2.5pt \hbox{$<$}}\lower 2.5pt
\hbox{$\sim$}}}

\newcommand{\Lumint}{{\cal L}_{\rm int}}

\begin{document}
\title{HARD BREMSSTRAHLUNG PHOTONS \\ FROM GRAVITON EXCHANGE\thanks{This 
work is supported in part by the Research Council of Norway.}
}

\author{T. BUANES, E.W.\ DVERGSNES
         \ and 
        P.\ OSLAND\thanks{e-mail addresses: 
Buanes@ift.uib.no, Erik.Dvergsnes@ift.uib.no, Per.Osland@fi.uib.no}
\\
\\
{\it Department of Physics and Technology, University of Bergen,}
\\
{\it N-5007 Bergen, Norway}
}
\date{}
\maketitle
\begin{abstract}
We review photon Bremsstrahlung in extra-dimensional models 
with massive gravitons.
The photon spectrum is harder than in the Standard Model.
In the RS scenario, radiative return to gravitons below the c.m.\ energy
can lead to a considerable increase of the cross section.
\end{abstract}
\section{Introduction}
In popular models involving extra dimensions \cite{Arkani-Hamed:1998rs,
Randall:1999ee}, the coupling of gravitons to ordinary matter can result in
extraordinary phenomena. These effects lead to events with energy-momentum
imbalance due to graviton emission, as well as effects due to virtual graviton
exchange \cite{Giudice:1999ck,Hewett:1998sn,Davoudiasl:1999jd,Allanach:2000nr}.

The Randall--Sundrum (RS) model \cite{Randall:1999ee} predicts the existence
of TeV-scale gravitons, whose couplings to Standard-Model (SM) fields have
weak-interaction strength. They will therefore show up as narrow resonances
\cite{Davoudiasl:1999jd}.
We shall here report on a study of the effects of graviton exchange on the
process \cite{Buanes:2004ya}
\begin{equation} \label{Eq:eemumuga}
e^-(k_1) + e^+(k_2) \to \mu^-(p_1) + \mu^+(p_2) + \gamma(k),
\end{equation}
focusing in particular on the emitted photon.

The coupling of gravitons to fermions and photons lead to additional
Feynman diagrams, beyond those which have the topology of the SM diagrams,
where the photon or $Z$ is replaced by a graviton. The additional diagrams
arise since the graviton also couples to two photons, and since there is
a fermion-fermion-vector-graviton coupling \cite{Giudice:1999ck,Han:1999sg}.

With the notation of (\ref{Eq:eemumuga}) the graviton propagator will be
characterized by either of two invariants: $s=(k_1+k_2)^2=4E^2$, or
$s_3=(p_1+p_2)^2$, and we classify the Feynman diagrams accordingly. The
terminology ``ISR'' and ``FSR'' will thus refer to whether the photon is
emitted ``before'' or ``after'' the graviton that is exchanged
\cite{Buanes:2004ya}.

Since some of the ``new'' Feynman diagrams do not have the collinear and IR
singularities of the SM contributions, one expects the photons to be harder
and less aligned with the initial or final-state fermions. This turns indeed
out to be the case in the ADD scenario.  Another striking feature in the RS
scenario is that radiative return can lead to peaks in the cross section
related to resonant production of gravitons whose masses are below the c.m.\
energy.
\section{Cross sections}
The differential cross section for (\ref{Eq:eemumuga}) is given in
\cite{Buanes:2004ya} in terms of the fractional energies of the muons ($x_1$,
$x_2$) and the photon ($x_3$) and the angle ($\theta$) between the photon and
the initial beam. We shall here review integrated cross sections, and some
photon $k_\perp$ distributions, where ``perpendicular'' refers to
perpendicular w.r.t.\ the beam direction.

The total cross section is defined as:
\begin{equation}
\sigma_{ee \to \mu\mu\gamma}
=\int_{-1+c_\text{cut}}^{1-c_\text{cut}} d(\cos\theta) 
\int_{x_3^\text{min}}^{x_3^\text{max}} dx_3
\int_{-x_3 + y_\text{cut}}^{x_3 - y_\text{cut}} d\eta
\frac{d^3\sigma_{ee \to \mu\mu\gamma}}{dx_3 d\eta\,d(\cos\theta)},
\end{equation}
where $\eta=x_1-x_2$ and
and we impose the cuts
$|\cos\theta|<1-c_\text{cut}$, with $c_\text{cut}=0.005$,
$x_3^\text{min}<x_3<\half(1-y_\text{cut}) \equiv x_3^\text{max}$,
$x_3^\text{min}=\max(\xi_\text{cut}/\sqrt{1-\cos^2\theta},y_\text{cut})$,
with $k_\perp^\text{min}=\xi_\text{cut}\sqrt{s}$, and
$y_\text{cut}=\xi_\text{cut}=0.005$.
In order to enhance the effects of graviton exchange, we consider a cut on
radiative return to the $Z$:
$s_3>(m_Z+3\Gamma_Z)^2 \equiv y_\text{cut}^\text{rr} s$.
\section{The RS scenario}
The discrete mass values of the RS scenario are given by
$m_n = (x_n/x_1) m_1$, $J_1(x_n)=0$,
with $J_1(x)$ a Bessel function. 
The Kaluza--Klein graviton coupling is given by 
$\kappa = \sqrt{2} (x_1/m_1) (k/\overline M_\text{Pl})$.
We parametrize the coupling strength by
$k/\overline M_\text{Pl}$, for which we consider three values: 0.01, 0.05, 0.1.
For a range of mass values $m_1$ assumed to
be of interest for a linear collider
\cite{Aguilar-Saavedra:2001rg}, we show in
Fig.~\ref{Fig:rs-masses-sigma} the masses of the next excited states.

\begin{figure}[htb]
\refstepcounter{figure}
\label{Fig:rs-masses-sigma}
\addtocounter{figure}{-1}
\begin{center}
\setlength{\unitlength}{1cm}
\begin{picture}(15,5.6)
\put(-1.0,.0)
{\mbox{\epsfysize=6.5cm\epsffile{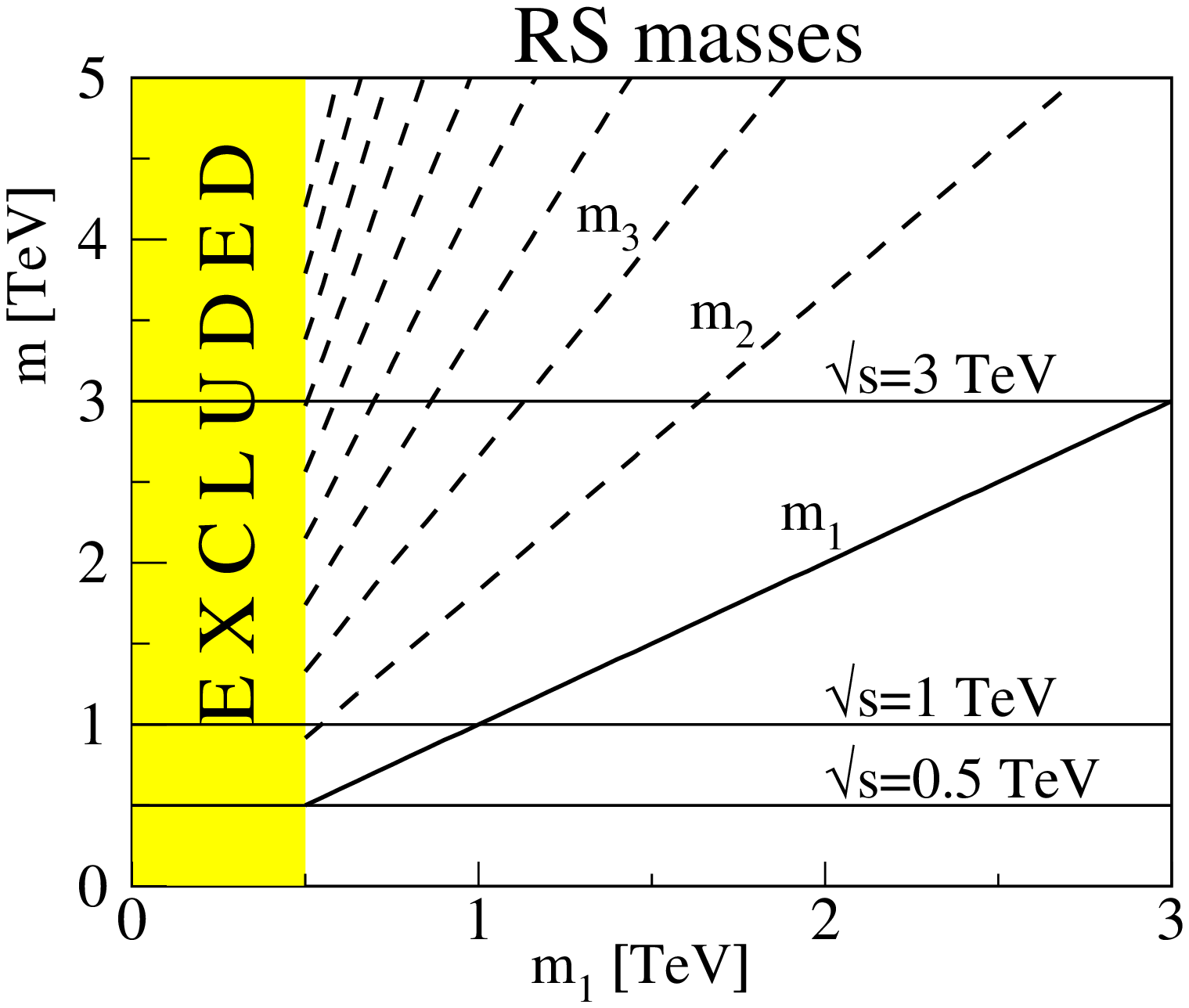}}
 \mbox{\epsfysize=6.5cm\epsffile{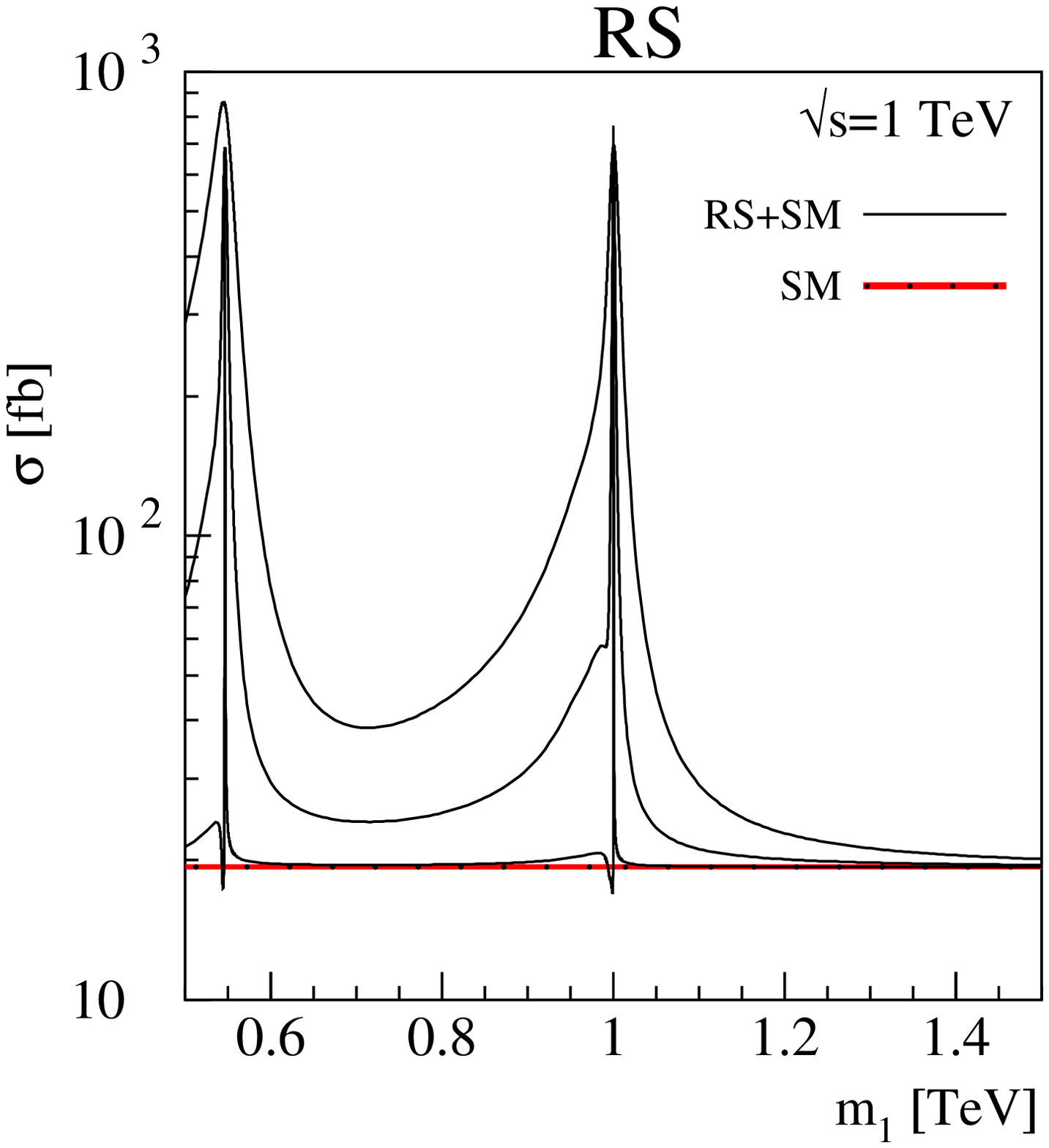}}}
\end{picture}
\vspace*{-12mm}
\caption{Left: Masses of RS gravitons vs.\ $m_1$.
Right:
Total cross sections for $e^+e^-\to\mu^+\mu^-\gamma$ vs.\ $m_1$, for
$\sqrt{s}=1$~TeV.
Three values of $k/\overline M_\text{Pl}$ are considered; 
from top and down: 0.1, 0.05 and 0.01.}
\end{center}
\vspace*{-5mm}
\end{figure}

If some RS graviton has a mass below the total c.m.\ energy, then an ISR
photon can carry away the right amount of energy to yield a resonance in the
$s_3$ channel. This effect of radiative return can give a significant
enhancement of the cross section as shown in Fig.~\ref{Fig:rs-masses-sigma}
for values of $m_1<\sqrt{s}$, in particular when the coupling is large.

\begin{figure}[htb]
\refstepcounter{figure}
\label{Fig:rs-dsigma-dkt}
\addtocounter{figure}{-1}
\begin{center}
\setlength{\unitlength}{1cm}
\begin{picture}(15,5.8)
\put(-0.5,0)
{\mbox{\epsfysize=6.2cm\epsffile{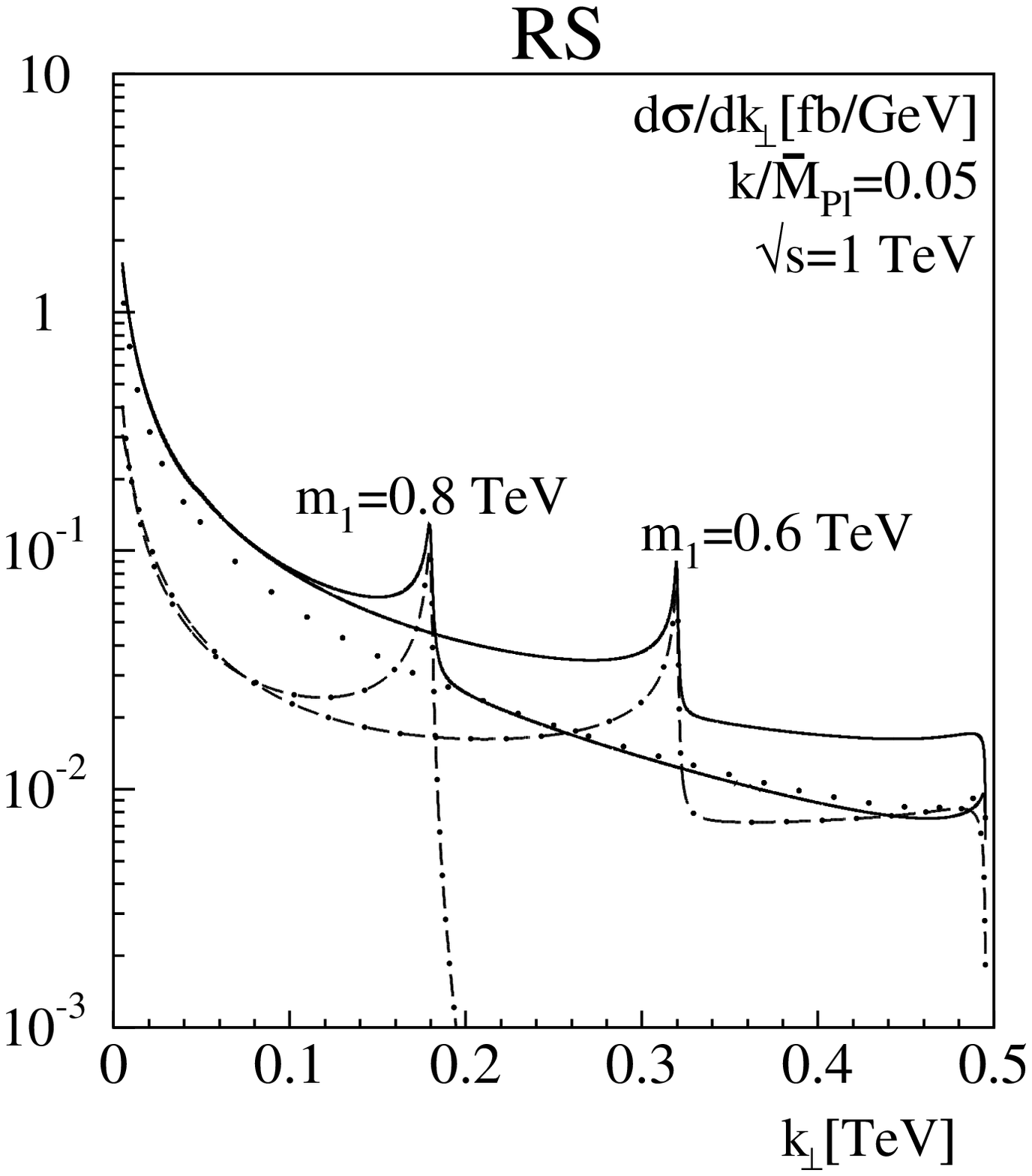}}
 \mbox{\epsfysize=6.2cm\epsffile{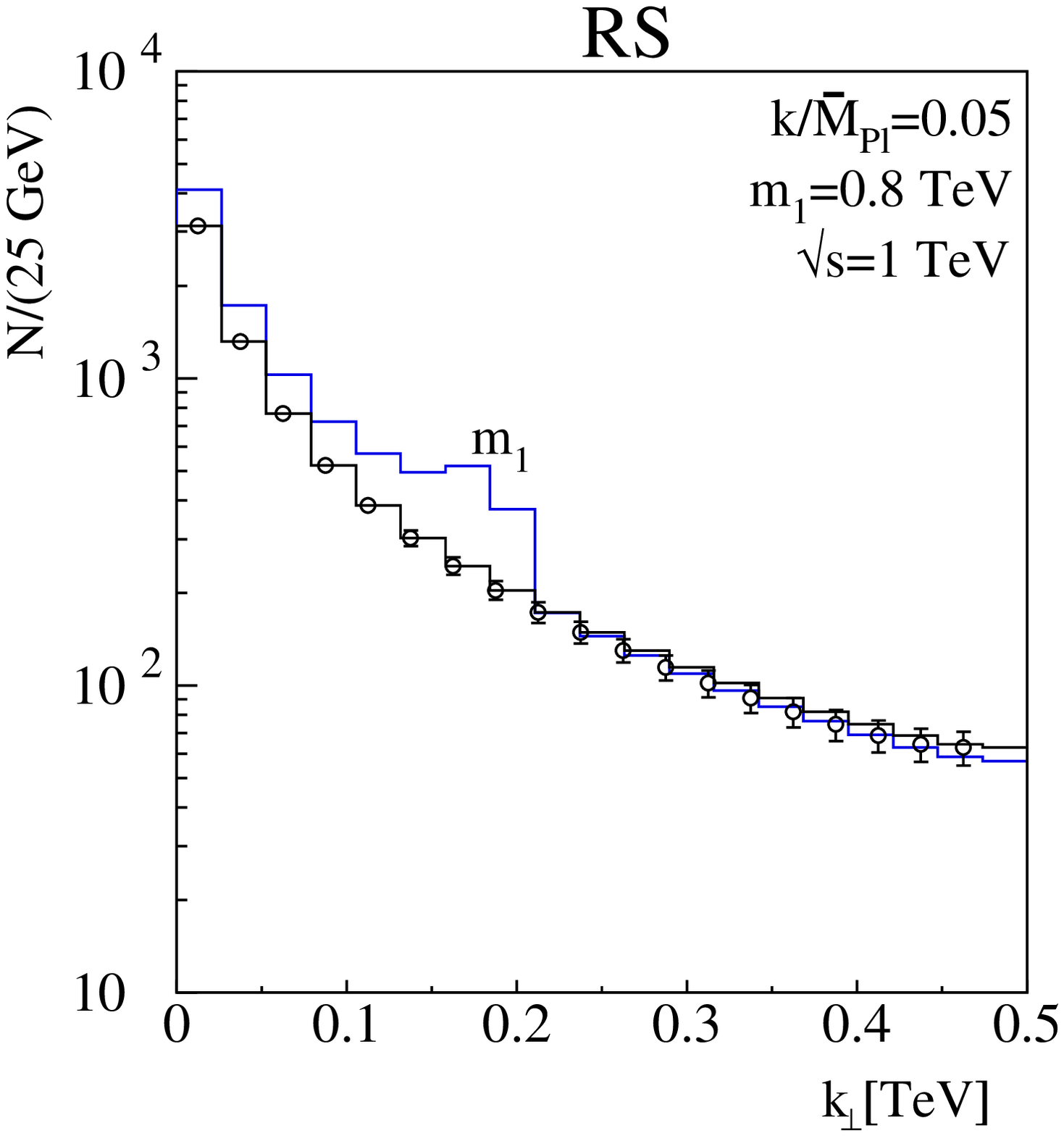}}}
\end{picture}
\vspace*{-10mm}
\caption{
Left: Photon perpendicular momentum distribution: solid, for
$m_1=0.6$ and 0.8~TeV.  
Graviton-related contributions: dash-dotted, SM contribution: dotted.
Right: bin-integrated $k_\perp$ distribution for
$m_1=0.8~\text{TeV}$.  Error bars (SM distribution)
for $\Lumint=300~\text{fb}^{-1}$. }
\end{center}
\vspace*{-5mm}
\end{figure}

The radiative return to lower-lying gravitons is even more clear in the photon
spectrum. In Fig.~\ref{Fig:rs-dsigma-dkt} we show the photon
perpendicular-momentum distribution for two values of $m_1<\sqrt{s}$.
Sharp peaks are seen for
$k_\perp\lsim(s-m_i^2)/2\sqrt{s}$.
(Note, however, that radiative return to the $Z$ has been excluded.)
The cross section at high $k_\perp$ is larger for $m_1=0.6$~TeV 
than for $m_1=0.8~\text{TeV}$, since (for $m_1=0.6$~TeV) $m_2=1.1$~TeV 
is close to $\sqrt{s}$.

\section{Summary}

The Bremsstrahlung process $e^+e^-\to\mu^+\mu^-\gamma$ may provide
confirmation of graviton exchange in two-body processes.  The photon spectrum
is harder, because of additional Feynman diagrams.  In the RS scenario, there
is an enhancement of the cross section due to radiative return.  Furthermore,
the angular distribution is enhanced at $\theta\sim\pi/2$
\cite{Buanes:2004ya}.

A related study on $e^+e^-\to\nu\bar\nu\gamma$ \cite{KumarRai:2003kk} also
finds peaks in the photon spectrum that will help identify the RS scenario.
This is a somewhat different signature, with a single photon plus missing
enery in the final state, but the underlying physics is analogous to what is
considered here.

At the LHC, the corresponding mechanism also leads to similar signatures.  In
the ADD scenario, there are hard photons and in the RS scenario there are
``steps'' in the $k_\perp$ distribution caused by radiative return
\cite{Dvergsnes:2002nc}.

\end{document}